\documentclass[11pt, letterpaper, final]{IEEEtran}

\usepackage{amsmath,amscd,amssymb}
\usepackage{bm}
\usepackage{graphicx}
\usepackage[tight]{subfigure}
\usepackage{units}
\usepackage{array,booktabs}
\usepackage{amssymb}
\usepackage{setspace}

\begin{document}

\title{Exact computations of the acoustic radiation force on a sphere using  the translational addition theorem}
\author{Glauber T. Silva,~\IEEEmembership{Member,~IEEE}, Andr\'e L. Baggio, J. Henrique Lopes, and Farid G. Mitri,~\IEEEmembership{Member,~IEEE} 

\thanks{Glauber T. Silva, Andr\'e L. Baggio, and  J. Henrique Lopes  are with Physical Acoustics Group, Instituto de F\'isica, Universidade Federal de Alagoas, Macei\'o, AL 57072-900, Brazil. Email: \texttt{glauber@pq.cnpq.br} (see http://www.if.ufal.br/$\sim$gaf) }
\thanks{Farid G. Mitri is with Los Alamos National Laboratory, 
Materials Physics and Applications Division, 
MPA-11, Sensors \& Electrochemical Devices, 
Acoustics \& Sensors Technology Team, MS D429, 
Los Alamos, NM, USA 87545.}}

\maketitle

\begin{abstract}
In this paper, the translational addition theorem for spherical functions is employed to exactly calculate 
the acoustic radiation force produced by an arbitrary shaped beam  on a sphere suspended in an inviscid fluid.
The radiation force is given in terms of the beam-shape and the scattering coefficients.
Each beam-shape coefficient (BSC) is the complex weight of a multipole mode in partial-wave expansion of the incident beam.
Moreover, they depend on the choice of the reference frame which is defined by the sphere's center. 
On the other hand, the scattering coefficients are obtained from the acoustic boundary conditions across the sphere's surface.
Given a set of known BSCs, the translational addition theorem can be used to obtain 
the new coefficients relative to the sphere's position.
Such approach is particularly useful when no closed-form expression of the incident pressure is known,
but  its BSCs are available.
This is the case of a spherically focused ultrasound beam, for which we compute
the radiation force on an absorbing compressible sphere arbitrarly placed in the fluid.
The analysis is carried out in  the Rayleigh  and the resonant scattering regimes.  
It is shown that the focused beam may  trap a spherical particle  in the Rayleigh  scattering regime.
\end{abstract}


\section{Introduction}
An increasing interest on ultrasound radiation force has been trigged after the concept
of acoustical tweezers was introduced by Wu~\cite{wu:2140}.
In acoustophoresis, the radiation force of a ultrasound standing wave is used
in contactless manipulation, 
separation, and trapping of small particles and cells~\cite{lenshof:1210}.
Two-dimensional tweezers are achieved in lab-on-a-chip devices~\cite{shi:2890}
and by employing circular phased-arrays~\cite{courtney:195}.
Single-beam acoustical tweezer has also been developed by using 
a tightly focused beam~\cite{lee:073701}.

The acoustic radiation force exerted by a plane or a spherical wave on a suspended sphere in a 
nonviscous fluid has been extensively investigated over the last 
century~\cite{king:212, embleton40, yosioka:167, westervelt:26,gorkov:773, nyborg:947, hasegawa:1139}.
Based on the partial-wave expansion 
of the incident and the scattered waves, the axial radiation force exerted on a sphere
by a planar piston beam~\cite{hasegawa:1578,mitri:114102}, a spherically 
focused beam~\cite{chen:713}, Bessel beams~\cite{marston:3518, mitri:2840, mitri:1059,azarpeyvand:4337}, 
and  a Gaussian~\cite{zhang:2007} beam.
In all these studies, the sphere is located in the axis of the beam.
For focused beams, the analysis was restricted to the sphere placed in the
focus point.

The design of single-beam acoustical tweezers requires an analysis of how radiation force
on the sphere behaves in the vicinity of the transducer's focus point.
So far, this analysis has been only performed for the  geometric scattering 
regime~\cite{lee:1084}, for which $ka\gg1$, where $k$ is the incident wavenumber
and $a$ is the sphere's radius.
For other limits such as the Rayleigh ($ka\ll1$) and the resonant ($ka\sim 1$) scattering regimes,
the radiation force can be computed using the partial-wave expansion method~\cite{silva:3541}.
In this method, the radiation force exerted on a suspended sphere  
is expressed in terms of the beam-shape and the scattering coefficients~\cite{silva:298}.
Each beam-shape coefficient (BSC) is the complex amplitude of an incident partial-wave,
while the scattering coefficients are obtained from the acoustic boundary conditions across
the sphere's surface.
The BSCs carry information regarding to the geometry of the incident beam.
Similarly, the scattering coefficients is related to the mechanical properties of the scatter. 

The radiation force produced by a spherically focused transducer on a sphere arbitrarly located
 in the host medium can be obtained
by calculating the BSC with respect to the particle's position.
Numerical schemes to compute the BSC include the 
midpoint integral rule~\cite{silva:298,mitri:392} and the discrete spherical harmonics 
transform (DSHT)~\cite{silva:54003,silva:1207}, which is based on 
the discrete Fourier transform.
A description of other quadrature methods to compute BSCs in the context
of optical scattering is provided in Ref.~\cite{gouesbt:1537}.
Both methods require that  the incident pressure  amplitude
should be sampled over a virtual a sphere which encloses the beam propagation region,
which contains the spherical target.
For highly oscillating functions, the midpoint rule requires a large number of sampling points 
to ensure proper converge of the BSC computation.
Yet the DSHT renders more accurate results with less 
sampling points, it may develop numerical errors related to aliasing due to undersampling and 
 spectral leakage caused by function domain truncation.
 
In order to circumvent numerical approximations, we 
propose a 
method to exactly calculate the acoustic radiation force based on the partial-wave expansion
method~\cite{silva:3541, silva:1207} 
and the translational addition theorem~\cite{martin:book}.
A similar method has been divised to calculated the radiation pressure generated by an electromagnetic
wave~\cite{stout:1620}.
The proposed method is used to calculate the radiation force produced by a 
spherically focused transducer on a silicone-oil droplet.
The incident beam is generated considering the parameters of a typical 
biomedical focused 
transducer with an F-number of 1.6 and a driving frequency of $\unit[3.1]{MHz}$.
Using closed-form expressions of the BSCs with respect to the beam focal point~\cite{edwards:1006}, 
the radiation force is computed along the beam's axis
and on the transducer's focal plane.
Both  Rayleigh and resonant scattering regimes are considered in this analysis.
The results obtained for the Rayleigh regime are compared to those
computed from Gorkov's theory~\cite{gorkov:773}.
A significant deviation from Gorkov's theory is noted in the axial radiation force 
when ultrasound absorption inside the droplet is 
taken into account.
In addition, transverse trapping is achieved in both Rayleigh and resonant scattering regimes.
Nevertheless, simultaneous axial and transverse trapping only occurs for droplets in the Rayleigh scattering regime.

\section{Physical model}
Consider an acoustic beam of arbitrary wavefront with angular frequency $\omega$ 
that propagates in an inviscid infinite fluid.
The fluid has ambient density $\rho_0$ and
speed of sound $c_0$.
The acoustic beam is described by the excess of pressure $p$  as a function of the position vector ${\bf r}$, with respect
to a defined coordinate system.
The time-dependence $e^{-i\omega t}$ is suppressed for the sake of simplicity.
An spherical scatterer with radius $a$, density $\rho_1$, and speed of sound $c_1$ is placed in the beam path.

\subsection{Scattering problem}
A spherically focused transducer
with diameter $2b$ and curvature radius $z_0$ is used to produce a focused beam  
(see Fig.~1) 
The origin of the coordinate system $O$ is set at the transducer's focus.
When the center of the sphere coincides with $O$,
the scattering of the incident beam 
is referred to as the on-focus scattering configuration~\cite{edwards:1006}.

In the on-focus scattering formalism, the normalized amplitude of the incident pressure beam can be described in 
spherical coordinates 
${\bf r} = r {\bf e}_r(\theta, \varphi)$, where ${\bf e}_r$ is the radial unit-vector,
$\theta$ and $\varphi$ are the polar and the azimuthal angles, respectively.
The incident pressure partial-wave expansion is given by~\cite{williams:book}
\begin{equation}
\label{pi}
 p_\text{i} = \sum_{n,m} a_n^m j_n(k r) Y_n^m(\theta, \varphi),
\end{equation}
where  $\sum_{n,m}=\sum_{n=0}^\infty\sum_{n=-m}^m$, $a_n^m$ are the BSCs to be determined, 
$k=\omega/c_0$,
$j_n$ is the $n$th-order spherical Bessel function,
and $Y_n^m$ is the spherical harmonic function of $n$th-order and $m$th-degree.
Note that the amplitude is normalized to the pressure magnitude $p_0$.
Using the orthogonality properties of the spherical harmonics, the BSCs are obtained from ~(\ref{pi}) as
\begin{equation}
\label{anm}
a_n^m = \frac{1}{j_n(k R)}\int_\Omega p_\text{i}(kR,\theta,\varphi) Y_n^{m*}(\theta,\varphi) d\Omega,
\end{equation}
where $R$ is the radius of a virtual spherical region where the beam propagates, $d\Omega$ is the differential solid 
angle, and $\Omega=\{(\theta,\varphi): 0\le \theta \le \pi, 0 \le \varphi \le 2 \pi \}$.

Suppose now that the sphere is translated to a new point denoted by a vector $\bf d$  as shown in Fig.~1.
This corresponds to the off-focus scattering by the sphere.
In spherical coordinates, the translational vector $\bf d$ is represented 
with respect to the system $O$ by $(d,\theta_d, \varphi_d)$.
The sphere's center  defines a new coordinate system denoted by $O'$.
The incident beam can be described in a new spherical coordinate
system $(r',\theta',\varphi')$ as
\begin{equation}
\label{pi2}
 \hat{p}_\text{i} = \sum_{\nu,\mu} {a_\nu^\mu}' j_\nu(k r') Y_\nu^\mu(\theta', \varphi'),
\end{equation}
where ${a_\nu^\mu}'$ are called the translational BSCs.
A similar expression to ~(\ref{anm}) can be obtained for the translational BSCs
by using the orthogonal property of the spherical harmonics.

The relation between the position vectors of the $O$ and $O'$ coordinate systems 
is ${\bf r} = {\bf r}' + {\bf d}$.
The translational addition theorem between the wave functions
in the systems $O$ and $O'$ states that~\cite{martin:book}
\begin{equation}
\label{at}
 j_n(k r) Y_n^m(\theta,\varphi) = \sum_{\nu,\mu} S_{n\nu}^{m\mu}(k {\bf d})j_\nu(k r')
Y_\nu^\mu(\theta',\varphi'),
\end{equation}
where the separation matrix is given by
\begin{align}
\nonumber
  S_{\nu{n}}^{\mu{m}}(k{\bf d}) &= 
 4 \pi (-1)^{m} 
 \sum_{q=|n-\nu|}^{n+\nu} i^{q+n-\nu} \mathcal{G}(\nu,\mu; n, -m;q ) \\ 
&\times j_q(kd) Y_q^{\mu-m}(\theta_d, \varphi_d),
\label{S}
\end{align}
with  $\mathcal{G}$ being the Gaunt coefficient~\cite{martin:book}.
This coefficient is zero  if any of the following conditions happens $n+\nu+q$ is odd,
$q>\nu+n$, or $q<|n-\nu|$.

Now, substituting ~(\ref{at}) into ~(\ref{pi}) and use the result
in ~(\ref{anm}) one obtains the translational BSCs as
\begin{equation}
\label{bnm2}
 {a_\nu^\mu}' = \sum_{n,m} a_n^{m} S_{\nu n }^{\mu m}(k{\bf d}), 
\end{equation}
where $\nu = 0,1,\dots, \infty$ and $-\nu\le \mu \le \nu$.
Note that each translational BSC is a combination 
of all possible BSC with respect to the system $O$.
 \begin{figure}
  \includegraphics[scale=.5]{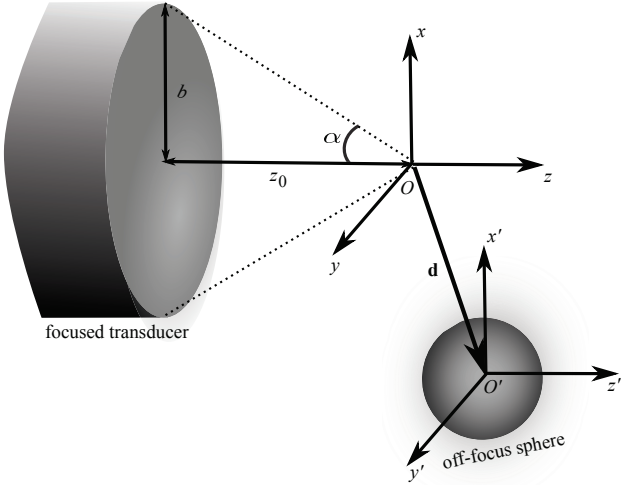}
 \caption{\label{fig:coords} Sketch of the acoustic scattering by a sphere
 placed anywhere in the host medium. 
 The systems $O$ and $O'$ correspond to the on-  and off-focus scattering
 configurations.}
 \end{figure}

Now, the scattered pressure is considered.
Assume that the sphere is
located at the transducer's focus  (i.e. on-focus scattering configuration).
The amplitude of the scattered pressure is described by the following 
expansion~\cite{williams:book}
\begin{equation}
p_\text{s} = \sum_{n,m} s_n^m h_n^{(1)}(k r)Y_n^m(\theta,\varphi),
\end{equation}
where $s_n^m$ is the scattering coefficient, $h_n^{(1)}$ is the $n$th-order spherical
Hankel function of first type.
Considering a compressional sphere, the scattering coefficient is determined from the continuity condition for
the pressure and the particle velocity across the sphere's surface.
Further details are provided in Ref.~\cite{anderson:426}.
Thus, using the aforementioned boundary conditions, one obtains
\begin{equation}
\label{snm}
s_n^m =  s_n a_n^m, 
\end{equation}
where 
\begin{align}
\nonumber
s_n &= -
\det
\left[
\begin{matrix}
 \gamma j_n(k a) & j_n(k_1 a)\\
j_n'(k a) & j_n'(k_1 a)
\end{matrix}
\right] \\
&\times \det
\left[
\begin{matrix}
 \gamma h_n^{(1)}(k a) & j_n(k_1 a)\\
{h_n^{(1)}}'(k a) & j_n'(k_1 a)
\end{matrix}
\right]^{-1},
\end{align}
with
$\gamma= \rho_0 k_1  / (\rho_1 k) $.
The prime symbol indicates differentiation with respect to the argument. 
According to ~(\ref{snm}) the scattering coefficient depends on the BSCs with respect
to the system $O$ and on  the sphere's mechanical parameters $s_n$, which
is independent of the choice of the coordinate system.
Therefore, in the system $O'$ the scattering coefficient becomes
\begin{equation}
{s_n^m}' = s_n {a_n^m}'.
\end{equation}

Ultrasound absorption inside  the sphere is considered by
adding an imaginary part to the wavenumber as follows~\cite{szabo:491}
\begin{equation}
 k_1= \frac{\omega}{c_1} +  i\alpha_0,
\end{equation}
where $\alpha_0$ is the absorption coefficient.
It is remarked that shear wave propagation inside the sphere is neglected.

\subsection{Beam-shape coefficients of a focused beam}
The amplitude of the pressure generated by a transducer  is given in terms of the Rayleigh integral
by~\cite{pierce:book}
\begin{equation}
\label{rayleigh}
p_\text{i}= -\frac{i k \rho_0 c_0 v_0 }{2\pi}\int _{S}\frac{e^{ik|{\bf r} - {\bf r}' |}}{|{\bf r} - {\bf r}'|}dS',
\end{equation}
where $S$ is the surface of the transducer's active element and $v_0$ is the uniform velocity distribution 
on $S$.
Physically, the Rayleigh integral expresses the Huygens' principle
in which the pressure at a position ${\bf r}$ is the sum of wavelets generated on
the surface $S$.
Hereafter, the pressure amplitude is normalized to $p_0=\rho_0 c_0 v_0$.

To derive the on-focus BSCs,
the origin of the coordinate system is centered at the transducer focal point. 
By assuming that $r\ll z_0$, with  $\|{\bf r}'\| = z_0$, the following 
expansion is used~\cite{colton:book}
\begin{eqnarray}
\nonumber
 \frac{e^{ik|{\bf r} - {\bf r}' |}}{|{\bf r} - {\bf r}' |} &=& 4\pi \frac{e^{ik z_0}}{z_0} \sum_{n,m}(-i)^{n}
j_n(kr)\\
&\times & Y_n^m(\theta',\varphi') Y_n^{m*}(\theta,\varphi)
\end{eqnarray}
into ~(\ref{rayleigh}).
Thus, integrating in the angular variables $(\theta',\varphi')$ yields
\begin{align}
\nonumber
p_\text{i} &= -i k z_0 e^{ik z_0} \sum_{n=0}^\infty i^n\sqrt{\frac{4 \pi}{2n+1}}\\
&\times [P_{n+1}(\cos \alpha) - P_{n-1}(\cos \alpha)]j_n(kr) Y_n^0(\theta, \varphi),
\label{pi_focused}
\end{align}
where $\alpha=\sin^{-1}(b/r_0)$ is the half-spread angle of the transducer and $P_n$ is the Legendre polynomial 
 of order $n$.
The on-focus BSCs are found by comparing Eqs.~(\ref{pi}) and (\ref{pi_focused}).
Accordingly,
\begin{eqnarray}
\nonumber
a_n^m &=& -i^{(n+1)} k z_0 e^{ikz_0} \sqrt{\frac{4 \pi}{2n+1}}\\
  &\times& [P_{n+1}(\cos \alpha) - P_{n-1}(\cos \alpha)]
\delta_{m,0}.
\label{alm_focused}
\end{eqnarray}

The partial-wave expansion of the focused beam can be compared to results
based on the paraxial approximation in ~(\ref{rayleigh}).
In adopting this approximation, it is assumed that  $b^2 \ll z_0^2$.
Let the radial distance in cylindrical coordinates be given by 
$\varrho=\sqrt{x^2 + y^2}$.
The pressure produced by the transducer in the focal plane is given by~\cite{lucas:1289}
\begin{equation}
\label{pi_rayleigh}
 p_\text{i}(\varrho,0) = i  \left(\frac{b}{\varrho}\right) \exp\left[ik\left(\frac{\varrho^2}{z_0} +  z_0\right)\right]
 J_1(k \varrho \sin\alpha),
\end{equation}
where $J_1$ is the first-order Bessel function.
Along the transducer's axis, we have
\begin{equation}
 p_\text{i}(0,z) =  \frac{iz_0}{z - z_0} \left\{ 1 - \exp
\left[\frac{ikb^2}{2}\left(\frac{1}{z}-\frac{1}{z_0}\right) \right]\right\}.
\end{equation}
These equations will be compared to the result obtained by the partial-wave expansion given in ~(\ref{pi_focused}).

\section{Acoustic radiation force}
After calculating  both the incident and scattered acoustic fields through the 
partial-wave expansion method, 
the radiation force
can be calculated by integrating the radiation stress tensor ${\bf S}$ over the scatter's surface.
In second-order approximation, the time-averaged radiation stress tensor is given by~\cite{westervelt:26}
\begin{equation}
{\bf S} = 
\rho_0 \overline{{\bf v} {\bf v}} - \left( \frac{\rho_0\overline{v^2}}{2} - 
\frac{\overline{p^2}}{2 \rho_0 c_0^2}\right) {\bf I},
\end{equation}
where the overbar denotes time-average, $\rho_0{\bf v}{\bf v}$ is the
Reynolds' stress tensor, $v=\|{\bf v}\|$, and
${\bf I}$ is the $3\times3$-unit matrix.
Considering a nonviscous fluid, the radiation stress tensor is a zero divergent quantity, i.e. 
$\nabla \cdot {\bf S}=0$.
Thus, by using the Gauss divergence theorem one can show that the radiation force is given by integrating the radiation
stress tensor on a control spherical surface centered at the target object. 
Furthermore, the radius of the
control surface $r$ lies in the farfield, i.e. $kr\gg 1$.
Therefore, the radiation force is given by 
\begin{equation}
{\bf f} = - r^2 \int_{\Omega} {\bf S} \cdot {\bf e}_r d\Omega, \quad kr \gg 1.
\end{equation}
One can show that the Cartesian components of the radiation force can be expressed as~\cite{silva:3541,silva:1212}
\begin{equation}
\label{eq:force}
{\bf f}=\pi a^2E_0(Y_x{\bf e}_x+Y_y{\bf e}_y+Y_z{\bf e}_z),
\end{equation}
where $E_0=p_0^2/(2\rho_0c_0^2)$ is the characteristic energy density of the incident wave,
and ${\bf e}_x$, ${\bf e}_y$ and ${\bf e}_z$ are the Cartesian unit-vectors. 
The radiation force functions are given by
\begin{eqnarray}
\nonumber
Y_x + iY_y &=&\frac{i}{2\pi(ka)^2}\sum_{n,m}
 \sqrt{\frac{(n+m+1)(n+m+2)}{(2n+1)(2n+3)}} \\
&\times & \bigl( S_n a_n^m a_{n+1}^{m+1*}   + S_n^*  a_n^{-m*} a_{n+1}^{-m-1}  \bigr), 
\label{Yxy}
\\
\nonumber
Y_z&=& \frac{1}{\pi(ka)^2}\textrm{Im}\sum_{n,m}\sqrt{\frac{(n-m+1)(n+m+1)}{(2n+1)(2n+3)} }\\
&\times &  S_n a_n^m a_{n+1}^{m*},
\label{Yxyz}
\end{eqnarray}
where `Im' denotes the imaginary part, the symbol $^*$ means complex conjugation, and
\begin{equation}
\label{Sn}
S_n = s_n + s_{n+1}^* + 2 s_n s_{n+1}^*.
\end{equation}
It is worth to note that the radiation force functions $Y_x,Y_y,$ and $Y_z$ are real-valued quantities.

\section{Results and discussion}
Consider  an acoustic beam that is generated in water for which $c_0=\unit[1500]{m/s}$ and $\rho_0=\unit[1000]{kg/m^3}$.
The focused transducer has a radius $\unit[b=22]{mm}$, a F-number of $1.6$, and operates at $\unit[3.1]{MHz}$.
Note that $(b/z_0)^2 = 0.1$, which ensures the paraxial approximation for the incident beam.
The magnitude of the pressure generated by the transducer is $p_0=\rho_0 c_0 v_0 = \unit[10^5]{Pa}$.
A droplet  made out of silicone-oil ($c_1=\unit[974]{m/s}$, $\rho_1= \unit[1004]{kg/m^3}$, and $\alpha_0=\unit[21]{Np/m}$ 
at $\unit[3.1]{MHz}$)
is used as the target object.

The truncation error in computing the translational BSCs in ~(\ref{bnm2}) is analyzed
in Appendix A.
Accordingly, the truncation order for the index $n$ necessary to achieve a certain error
$\epsilon$ is given by
\begin{equation}
 N = \nu + k d + 1.8 (\log\epsilon^{-1})^{2/3} (kd)^{1/3}.
\end{equation}
The error considered in the translational BSC computations is $\epsilon=10^{-6}$.
The truncation order $L$ of the radiation force series in (\ref{Yxyz})
is established by the scattering coefficient ratio $|s_L^m/s_0^0|< 10^{-7}$.

In Fig.~2, 
the pressure amplitude (normalized to $p_0$) produced by the focused transducer is shown along 
both $x$ and $z$ directions.
The pressure is computed using the partial-wave expansion method  and 
the paraxial approximation based on 
 ~(\ref{pi_rayleigh}).
Good agreement is found between the methods in the transverse direction and in the 
vicinity of the focal region.
Though moving away from the focal region, the partial-wave expansion method deviates from the 
paraxial approximation  result.
This happens because the partial-wave expansion is valid in the vicinity of the focal region, i.e. $r\ll z_0$.
 \begin{figure}[h]
 \centering
  \includegraphics[scale=.5]{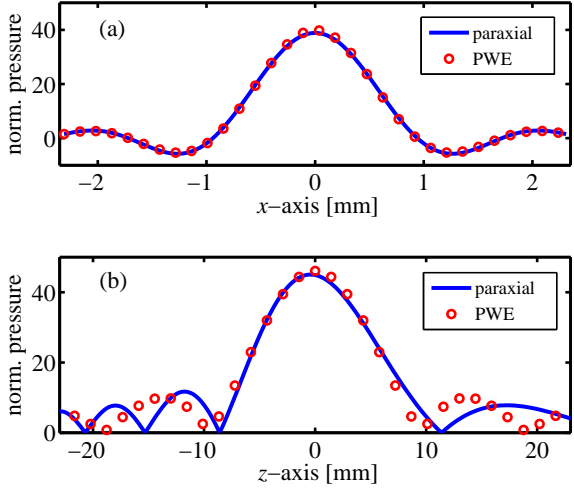}
 \caption{\label{fig:validation} Pressure amplitude generated by the spherically 
 focused transducer with 
 aperture $\unit[44]{mm}$ and F-number of $1.6$, operating  at $\unit[3.1]{MHz}$.
 The pressure is evaluated along (a) the transverse and (b) the axial directions
 using the partial-wave expansion (PWE) method and the paraxial approximation.}
 \end{figure}
 \begin{figure}[h!]
  \centering
  \includegraphics[scale=.5]{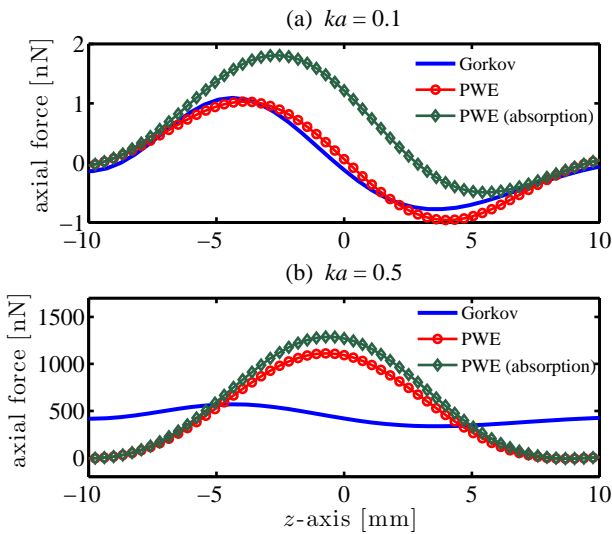}
 \caption{\label{fig:rad_force_z} 
 Axial radiation force versus the droplet's position along 
 $z$ direction in the focal plane.
 The force is computed through the partial-wave expansion (PWE) and Gorkov's methods.
 The size factors of the sphere are (a) $ka=0.1$  and (b) $ka=0.5$.
 }
 \end{figure}

Figure~3 
exhibits the axial radiation force versus  the droplet's position along the transducer's axis ($z$ direction).
The droplet's size factors are $ka=0.1$ (Rayleigh regime) and $ka=0.5$ (resonant regime).
The solid line represents the radiation force computed with Gorkov's 
method.
Good agreement between the methods is found when no ultrasound absorption
is considered in the droplet.
However, when ultrasound absorption is taking into account, a significant
deviation between these methods is observed.
This result is expected since the acoustic radiation force
caused by a plane traveling wave on a sphere depends on the sum of the scattered and absorbed power~\cite{westervelt:26}.
The Gorkov's theory does not take into account absorption inside the particle.
Therefore, if absorption is considered, the radiation force will be larger 
than in the case where absorption is neglected~\cite{lofstedt:2027,silva:pre}.
The droplet with  $ka=0.1$ is axially trapped along the transducer axis at 
$z=\unit[2.5]{mm}$ (with absorption).
Note that a nonabsorptive droplet would be trapped at $z=0$.
Gorkov's method no longer describes the behavior of the radiation force generated by the focused transducer when $ka=0.5$. 
Furthermore, it is not possible to axially trap the droplet with the analyzed transducer at this size factor.
This can be understood as follows.
The radiation force exerted on a small particle by the spherically focused beam is formed by two contributions~\cite{gorkov:773}: ``the scattering force'' caused mostly by the traveling wave part 
of the beam, while ``the gradient force'' is due to the spatial variation of the  potential and kinetic energy densities of the 
beam.
As the size factor increases, so does the scattering force.
When this force overcomes the gradient force, the axial
trapping is no longer possible.
This might be prevented using tightly focused beams~\cite{lee:073701}.

The transverse radiation force versus the droplet's position along
 $x$ direction in the focal plane is shown in Fig.~4. 
Excellent agreement is found between the partial-wave expansion and Gorkov's methods 
when $ka=0.1$.
Some deviation between the methods arises when $ka=0.5$.
Note that the results with and without absorption are very alike.
The transverse trapping happens for both $ka=0.1$ and $0.5$, because no scattering radiation force is present in the transverse direction.
Only the gradient radiation force appears in this direction.
Based on Figs.~3 
and 4
we  conclude that the focused transducer forms a 3D acoustical tweezer for silicone-oil droplets in the Rayleigh scattering regime.
 \begin{figure}[h]
  \centering
  \includegraphics[scale=.6]{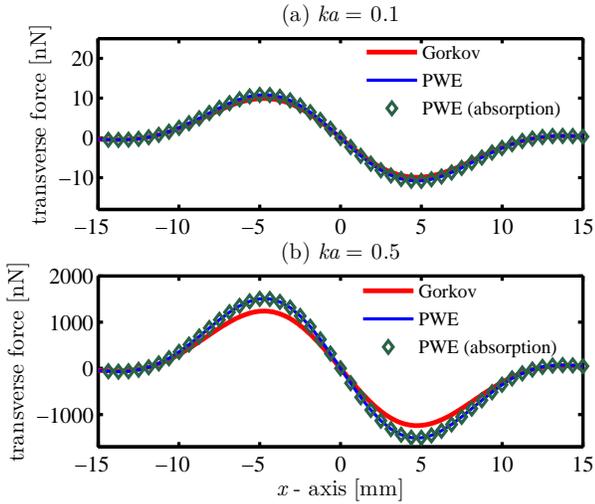}
 \caption{\label{fig:rad_force_trans} 
 Transverse radiation force versus the droplet's position along 
 $x$ direction in the focal plane.
 The force is computed through the partial-wave expansion and Gorkov's methods. 
 The size factors are (a) $ka=0.1$  and (b) $ka=0.5$.}
 \end{figure}

The axial radiation force versus the droplet's position along $z$ direction for
 the resonant scattering regime ($ka=1$ and $5$)
is shown in Fig.~5. 
In this case, the scattering force totally overcomes the gradient force.
Thereby, no trapping is possible in the axial direction with the analyzed transducer.
The axial radiation force pushes the droplet to the forward scattering direction.
 \begin{figure}
  \centering
  \includegraphics[scale=.6]{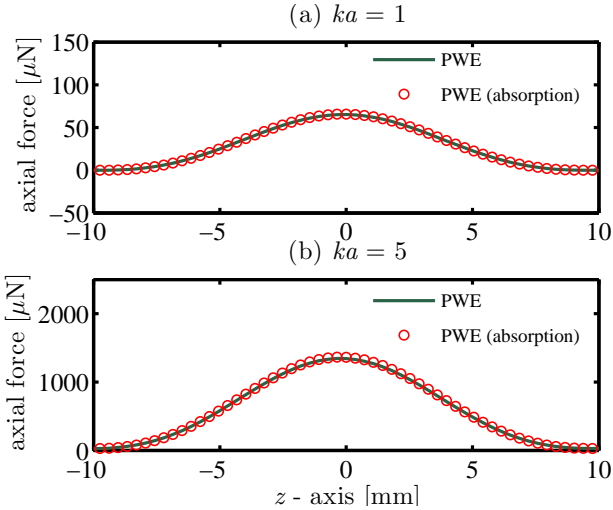}
 \caption{\label{fig:rad_force_z2} Axial radiation force versus the droplet's
 position along $z$ direction for
 the resonant scattering regime (a) $ka=1$ and (b) $ka=5$.}
 \end{figure}

In Fig.~6, 
the transverse radiation force versus the  droplet's
position along $x$ direction is displayed for the resonant scattering regime 
($ka = 1$ and $5$).
It is clearly shown that the transverse trapping is still possible in this scattering regime.
This happens because the incident beam is tightly focused in the transverse direction.
Moreover,  the transverse radiation force remains practically the same  
regardless of ultrasound absorption within the droplet.
 \begin{figure}[h]
  \centering
  \includegraphics[scale=.6]{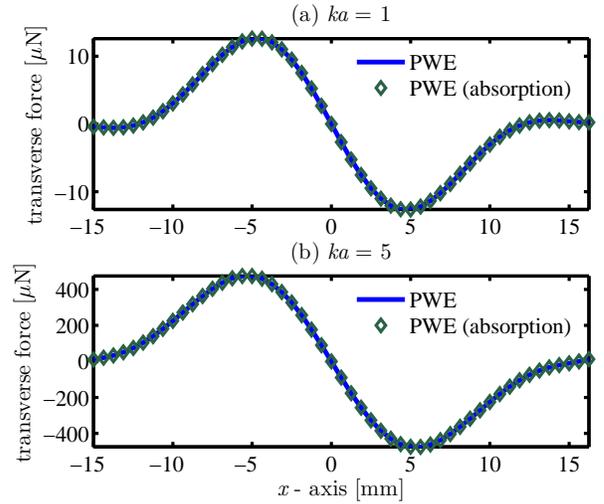}
 \caption{\label{fig:rad_force_trans2} Transverse radiation force on a silicone oil droplet 
 (with and without attenuation) in the resonant scattering regime (a) $ka=1$ and (b) $ka=5$.}
 \end{figure}

The vector field of the transverse radiation force on the 
silicone-oil droplet placed in
the transducer focal plane is displayed in Fig.~7. 
The background map represents the axial radiation force exerted on the droplet.
If the droplet lies in the circular region with radius of $\unit[0.5]{mm}$ around the 
focus point, it will be attracted and trapped along the transducer's axis. 
The droplet is transversely trapped by a force of about $\unit[10]{\mu N}$ (see Fig.~\ref{fig:rad_force_trans}),
but it 
will be further pushed axially by a force of  $\unit[60]{\mu N}$.
Therefore, the focused transducer operates as a 2D acoustical tweezer for droplets in 
the resonant scattering regime ($ka=1$).
 \begin{figure}[h]
 \begin{center}
  \includegraphics[scale=.6]{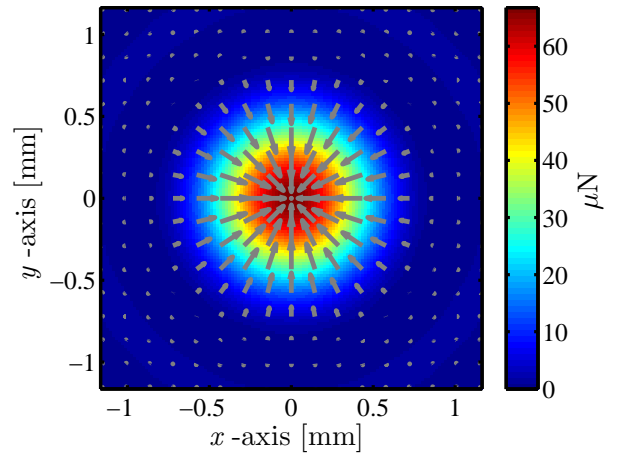}
 \end{center}
 \caption{\label{fig:quiver1} Vector field of the radiation force in the transducer focal plane
 produced on the silicone oil droplet in the resoant regime $ka=1$. 
 The vector field  is plot on top of the axial radiation force.}
 \end{figure}

\section{Summary and conclusions}
A method to compute the axial and the transverse acoustic radiation force 
on a sphere
based on the translational addition theorem for spherical wave functions
was presented.
The method relies on the fact that once the BSCs are known with respect to
a coordinate system, they can be calculated in a new  translated coordinate system.
The radiation force generated by an ultrasound 
focused beam in the paraxial approximation and exerted on a silicone-oil droplet
placed anywhere in the host medium was calculated using the proposed method.
Both axial and transverse radiation forces were computed in the Rayleigh and resonant scattering regimes.
In the Rayleigh regime, the obtained results were compared to Gorkov's radiation force theory.
Good agreement was found between these methods when ultrasound absorption inside the droplet was neglected.
Nevertheless, a significant deviation between the results  takes
place in the axial radiation force when the ultrasound absorption is considered.
It was showm that
the transducer under investigation, with driving frequency of $\unit[3.1]{MHz}$ and an F-number of 1.6, 
can operate as a single-beam acoustical tweezer in 3D for the Rayleigh
scattering regime.
In contrast, only transverse trapping of the silicone-oil droplet
was achieved for the resonant scattering regime. 

In conclusion, the translational addition theorem of spherical wave functions was used 
in combination with the partial-wave expansion to compute the axial and the transverse acoustic radiation force on a sphere. 
This method may become a useful tool in  the design and evaluation of acoustical tweezers
in particle manipulation applications.

\section*{Acknowledgements}
\label{Ack}
This work was supported by grants CAPES 2163/2009-AUX-PE-PNPD, CNPq 306697/2010-6,
and CNPq 481284/2012-5 (Brazilian agencies).

\appendix
\section{Truncation error}
Assume the series in ~(\ref{bnm2}) is truncated in $n=N$.
Thus, the truncation error is given by
\begin{equation}
\label{error}
\epsilon= \left|\sum_{n=N+1}^\infty\sum_{m=-n}^{n} a_{n}^{m} 
S_{\nu n }^{\mu m}(k{\bf d})
\right|.
\end{equation}
We consider the truncation error for a plane progressive wave.
This analysis as the upper limit case for the error because
the normalized pressure amplitude of an acoustic beam is smaller
than that of a normalized plane wave.
The BSC of a plane wave described by
$\hat{p}=e^{i k z}$ is $a_n^m=i^n \sqrt{4 \pi  (2n+1)} \delta_{m,0}$.
Substituting this into ~(\ref{error}) yields
\begin{equation}
\label{error2}
\epsilon =  \sqrt{4 \pi} \left|\sum_{n=N+1}^\infty i^n \sqrt{2n+1} S_{\nu, n }^{\mu, 0}(k{\bf d})
\right|.
\end{equation}
The leading term of the separation matrix in ~(\ref{S}) is related
to the spherical Bessel function of smallest order, i.e. $q=|n-\nu|$.
Therefore, the truncation error can be approximated to
\begin{align}
\nonumber
\epsilon \simeq  (4 \pi)^{3/2} \sqrt{2N-2\nu+3} \bigl|\mathcal{G}(\nu,\mu; N+1, 0;N-\nu+1 )\\
\times
j_{N-\nu+1}(kd) Y_{N-\nu+1}^{\mu}(\theta_d, \varphi_d)\bigr|.
\end{align}
After integrating over $(\theta_d, \varphi_d)$, we obtain
\begin{equation}
\epsilon\simeq  |j_{N-\nu+1}(kd)|.
\end{equation}
When the order of the spherical Bessel function is  larger than its argument, the error can be
approximated to~\cite{song:311}
\begin{align}
\nonumber
\epsilon \simeq \biggl| \frac{1}{2}\sqrt{\frac{1}{ x f(x,L)}}
\exp\biggl[
f(x,L)- \left(L + \frac{3}{2}\right)\\
\ln\biggl(L+\frac{3}{2} + \frac{f(x,L)}{x}
\biggr)
\biggr]\biggr|,
\end{align}
where $L=N-\nu +1$, $x=kd$, and $f(x,L) = \sqrt{(L+3/2)^2 - x^2}$.
We define $L+3/2 = x(1+\delta)$.
Since the spherical Bessel function decreases rapidly as the order becomes
larger than the argument, the parameter $\delta$ is assumed to be much smaller than the unit.
Therefore, the error can be expressed as
\begin{equation}
 \epsilon \simeq (2 \delta)^{-1/4} e^{-x(2\delta)^{3/2}/3}
\end{equation}
The second term in this equation is much smaller than the first and then dominates
the error.
Thus, taking the logarithm on both sides of this equation, we can estimate the truncation order
in ~(\ref{bnm2}) as
\begin{equation}
\label{N}
 N = \nu + k d + 1.8 (\log\epsilon^{-1})^{2/3} (kd)^{1/3}.
\end{equation}
The term $\log\epsilon^{-1}$ is closely related to the number of precision digits,
which is given by the nearest integer to $(\log\epsilon^{-1} + 1.0 - \log 2)$.


\end{document}